\documentclass[%
 reprint,
 superscriptaddress,
 amsmath,amssymb,
 aps,
prc,
]{revtex4-2}

\usepackage{graphicx}
\usepackage{dcolumn}
\usepackage{bm}
\usepackage{xcolor} 
\usepackage{hyperref}
\usepackage{float}
\usepackage{placeins}
\usepackage{ulem}
\usepackage{lineno}
\usepackage{booktabs}
\usepackage{soul}
\usepackage{multirow}

\begin{document}

\title{Reconstructability and directed flow of short-lived resonances in Au+Au collisions at 19.6 and 200 GeV}

\author{Junyi Han}
\affiliation{Key Laboratory of Quark and Lepton Physics (MOE) and Institute of Particle Physics, Central China Normal University, Wuhan 430079, China}

\author{Xialei Jiang}
\affiliation{Key Laboratory of Quark and Lepton Physics (MOE) and Institute of Particle Physics, Central China Normal University, Wuhan 430079, China}
\affiliation{GSI Helmholtzzentrum für Schwerionenforschung GmbH, Planckstr. 1, 64291 Darmstadt, Germany}

\author{Hongcan Li}
\affiliation{Key Laboratory of Quark and Lepton Physics (MOE) and Institute of Particle Physics, Central China Normal University, Wuhan 430079, China}

\author{Yaping Wang}
\thanks{Corresponding author: wangyaping@ccnu.edu.cn}
\affiliation{Key Laboratory of Quark and Lepton Physics (MOE) and Institute of Particle Physics, Central China Normal University, Wuhan 430079, China}

\date{\today}

\begin{abstract}
We present a systematic study of the reconstructability and directed flow of hadronic resonances in Au+Au collisions within the UrQMD transport model. The main objective of this work is to investigate how the hadronic stage influences both resonance reconstructability and the final-state directed flow.

A set of short-lived hadronic resonances, including $\rho^0$, $K^{*0}$, and $\Lambda(1520)$, is investigated to quantify their yields and reconstructable fractions as a function of charged-particle multiplicity, characterized by $(dN_{ch}/d\eta)^{1/3}$. 
We compare results at $\sqrt{s_{NN}} = 19.6$ and $200 ~\mathrm{GeV}$ to investigate possible energy-dependent differences in the reconstructability. Such differences reflect variations in the properties of the hadronic medium.
The results are further examined as a function of resonance lifetime, revealing a clear ordering of reconstructability among different resonances. Overall, the reconstructability is found to be primarily governed by resonance lifetime.
The directed-flow analysis reveals clear differences between resonances and their corresponding stable hadrons in mid-central collisions, while these differences become significantly weaker in peripheral collisions, highlighting the important role of hadronic evolution in shaping the final-state directed flow.
These studies provide a unified picture of how the hadronic stage influences both resonance reconstructability and directed flow, offering new insights into resonance observables in relativistic heavy-ion collisions.
\end{abstract}

\maketitle


\section{Introduction}
Relativistic heavy-ion collisions provide a unique opportunity to study strongly interacting QCD matter under extreme temperature and density. At sufficiently high energies, a deconfined quark-gluon plasma (QGP) is formed, which subsequently expands and cools, evolving into a hadron gas before kinetic freeze-out. While the properties of the QGP have been extensively studied, the later hadronic stage remains crucial for understanding how final-state hadrons are formed\cite{Shuryak1980, Heinz2013, STAR2005, PHENIX2005}.

During the hadronic phase, inelastic and pseudo-elastic interactions among hadrons can significantly modify final-state observables, including particle yields, correlations, and collective flow\cite{BraunMunzinger2003,CleymansSatz1993}. In particular, hadronic rescattering and regeneration processes play an important role in shaping the observable particle composition at kinetic freeze-out\cite{Bleicher2002,Markert2008}. Short-lived hadronic resonances are sensitive probes of the hadronic phase due to their lifetimes being comparable to the duration of the hadronic medium. A significant fraction of resonances decay inside the medium, and their decay daughters may undergo further rescattering, which destroys the invariant-mass correlation required for reconstruction. In addition, regeneration processes through hadronic interactions can recreate resonances via pseudo-elastic scattering. 
The measured resonance reflects both early-time dynamics and late-stage hadronic interactions\cite{Bass1998,Bleicher1999}.

Experimentally, a number of measurements of short-lived resonances have been performed in relativistic heavy-ion collisions, which indicate a general suppression of resonance yields in the hadronic phase.\cite{STAR_Kstar_2005,STAR_Kstar_2011,STAR_Kstar_BES_2023,STAR_rho0_2004,STAR_rho_f0_2004,STAR_Lambda1520_Sigma1385_2004,ALICE_Kstar_2024,ALICE_rho0_UPC_2021,ALICE_Lambda1520_ICHEP2022,TorrieriRafelski2001}. Motivated by these observations, the present work aims to systematically investigate how the hadronic phase modifies resonance production and related observables. In particular, we investigate the dependence on system size, collision energy, and resonance lifetime to quantify the effects of hadronic interactions.

For short-lived resonances such as $\rho^{0}$, $K^{*0}$, and $\Lambda(1520)$, a significant fraction of particles decay during the early stage of the hadronic evolution. In transport model simulations, these resonances can be reconstructed via their decay daughters using the similar invariant-mass technique as in experiments\cite{ALICEResonanceMethod2013,MPDResonanceReconstruction2021}. In addition to the reconstructed yields, we also use Monte Carlo truth-level resonances, defined as all generated particles that have decayed according to their natural decay widths. The reconstructability is defined as the fraction of resonances whose decay daughters do not undergo further interactions in the hadronic phase, and thus can be reconstructed via invariant-mass analysis\cite{Sahoo2024KstarRescattering}.
We then study the reconstructability as a function of charged-particle multiplicity, $(dN_{\mathrm{ch}}/d\eta)^{1/3}$, which serves as a proxy for the system size and medium density. This allows us to investigate the dependence of the reconstructability on the system size and medium properties, which directly reflects the strength of the interactions between decay daughters and the surrounding hadronic medium.

Owing to their different lifetimes—approximately $\sim 1.3$ fm/$c$ for $\rho^{0}$, $\sim 4$ fm/$c$ for $K^{*0}$, and $\sim 12$ fm/$c$ for $\Lambda(1520)$\cite{PDG2024_PRD}—these resonances probe different stages of the hadronic evolution. In particular, shorter-lived resonances decay earlier within the hadronic medium, and their decay daughters are therefore more likely to undergo rescattering with surrounding hadrons. This leads to a reduced probability of reconstructing the parent resonance via invariant-mass analysis. Consequently, the reconstructability is expected to exhibit a clear dependence on the resonance lifetime, which is systematically studied in this work.
We systematically compare the results between $\sqrt{s_{NN}}=19.6$ GeV and $\sqrt{s_{NN}}=200$ GeV collisions to investigate the energy dependence of hadronic interaction effects.

Additionally, we investigate the directed flow\cite{Ollitrault1992Flow} of resonances at different stages of the hadronic evolution, providing complementary information on the dynamical evolution of the hadronic medium. 
Together, these observables provide complementary probes of the hadronic stage, enabling a more comprehensive understanding of resonance evolution in relativistic heavy-ion collisions.

This paper is organized as follows. In Sec.~II, we describe the framework of the analysis and introduce the transport model setup. In Sec.~III, we present the results on resonance production and discuss their energy and lifetime dependence, as well as the direct flow at different stages of the hadronic evolution. Finally, a summary of the work is given in Sec.~IV.

\section{Framework}

We employ the UrQMD (Ultra-relativistic Quantum Molecular Dynamics) transport model to simulate Au+Au collisions at $\sqrt{s_{NN}} = 19.6$ and $200~\mathrm{GeV}$. UrQMD is a microscopic hadronic transport approach based on covariant propagation of hadrons in phase space and stochastic binary collisions, providing a unified description of particle production and hadronic interactions in relativistic heavy-ion collisions~\cite{Bass1998,Bleicher1999}. In this study, the cascade mode of UrQMD v4.0 is used.

In UrQMD, resonances are dynamically produced through hadronic interactions and string excitation and fragmentation processes, and subsequently decay according to their physical lifetimes and branching ratios.
To quantify hadronic medium effects, we distinguish between all resonance decays recorded in the event history and reconstructable resonances. The former are obtained from the event record (f16) and decay according to their physical widths, while the latter are defined as resonances whose decay daughters do not interact further in the hadronic phase.

In practice, reconstructable resonances are identified using the CTO(68) option in UrQMD v4.0, which enables the selection of resonances based on the daughter-particle rescattering information\cite{UrQMD_manual}. 
The reconstructability is defined as:
\begin{eqnarray}
R = \frac{N_{\mathrm{reconstructed}}}{N_{\mathrm{true}}}
\end{eqnarray}
where $N_{\mathrm{true}}$ denotes the total number of resonances that have decayed during the transport evolution, as obtained from the f16 event record, and $N_{\mathrm{reconstructed}}$ corresponds to those resonances whose decay daughters remain free from further rescattering. This observable directly quantifies the survival probability of resonances in the hadronic medium.

\section{Results}

\begin{figure}[htbp]
    \centering
    \includegraphics[width=1\columnwidth]{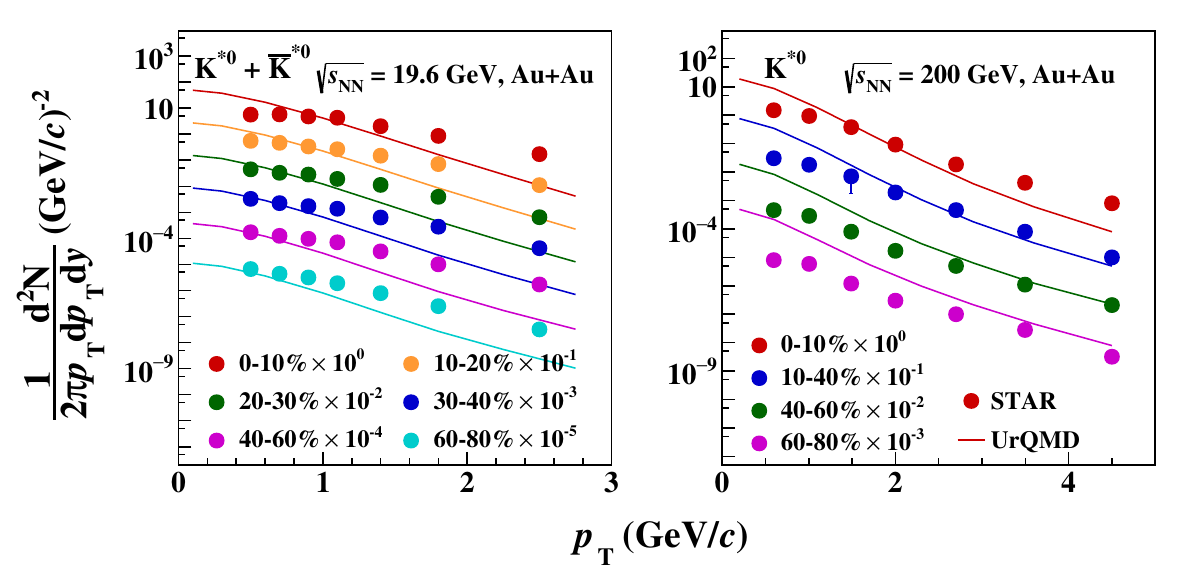}
    \caption{Transverse momentum $(p_T)$ spectra of $K^{*0} + \overline{K}^{*0}$ in Au+Au collisions at $\sqrt{s_{NN}} = 19.6~\mathrm{GeV}$ (left panel) and $K^{*0}$ at $\sqrt{s_{NN}} = 200~\mathrm{GeV}$ (right panel) from UrQMD simulations. Different centrality classes are shown with different colors as indicated in the legends. The spectra are scaled by successive powers of ten for clarity. The points represent STAR experimental data, while solid curves correspond to UrQMD model calculations.}
    \label{fig:fig1_pt}
\end{figure}

We first present the transverse momentum $(p_T)$ spectra of $K^{*0}$ production in Au+Au collisions at $\sqrt{s_{NN}} = 19.6$ and $200 ~\mathrm{GeV}$, as shown in Fig.~\ref{fig:fig1_pt}. The left panel shows the results for 19.6 GeV, where both $K^{*0}$ and $\overline{K}^{*0}$ are included, while the right panel corresponds to $200 ~\mathrm{GeV}$ where only $K^{*0}$ is shown for consistency with the available experimental results\cite{STAR_Kstar_BES_2023,STAR_Kstar_2005}.

The transverse momentum spectra are shown for different centrality classes and exhibit the expected steeply falling behavior with increasing transverse momentum. Different centrality classes are represented by different color markers, and successive scaling factors are applied for clarity. The UrQMD calculations (solid curves) are compared with available STAR data (symbols), showing a good description of the overall spectral shapes.

This comparison provides a baseline validation of the model performance for strange resonance production and establishes the reliability of the subsequent analysis of resonance-to-stable hadron ratios and reconstructability.

\subsection{Reconstructability}
\begin{figure}[htbp]
    \centering
    \includegraphics[width=1.0\columnwidth]{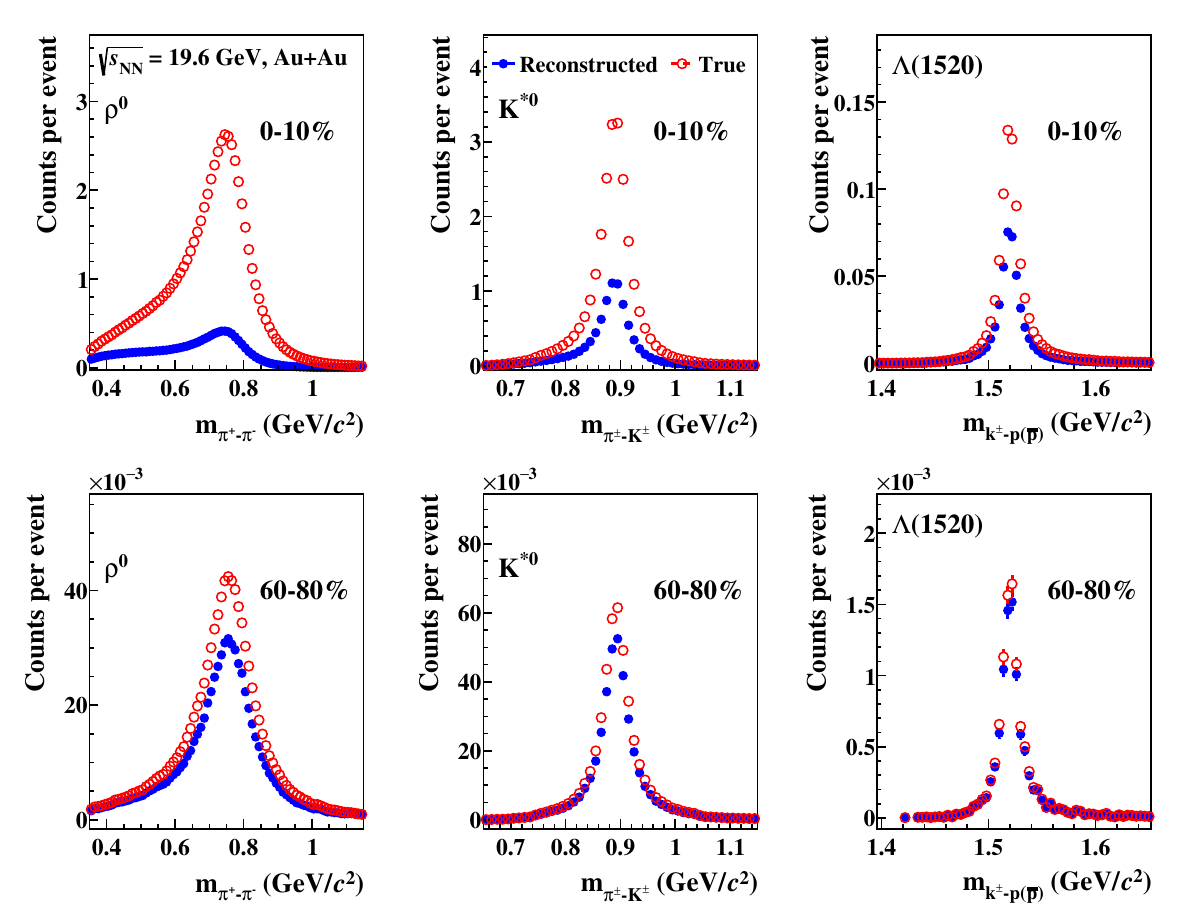}
    \caption{Invariant mass distributions for $\rho^{0}\rightarrow\pi^{+}\pi^{-}$, $K^{*0}\rightarrow\pi^{\pm}K^{\mp}$, and $\Lambda(1520)\rightarrow pK^{-}(\bar{p}K^{+})$ in Au+Au collisions at $\sqrt{s_{NN}}=19.6$~GeV for two centrality classes, $0-10\%$ (upper panels) and $60-80\%$ (lower panels). The reconstructed signals (blue points) are compared to the total number of particles that have actually decayed (red circles). The comparison demonstrates the performance of the reconstruction procedure for different collision centralities and resonance species.}
    \label{fig:fig2_invmass}
\end{figure}

We next examine the invariant mass distributions for $\rho^0 \rightarrow \pi^{+}+\pi^{-}$, $K^{*0} \rightarrow \pi^{\pm}+K^{\mp}$, and $\Lambda(1520) \rightarrow p+K^{-}$($\bar{p}+K^{+}$) in Au+Au collisions at $\sqrt{s_{NN}} = 19.6~\mathrm{GeV}$, as shown in Fig.~\ref{fig:fig2_invmass} for two centrality classes, $0-10\%$ and $60-80\%$. The reconstructed signals (blue points) are obtained using the CTO(68) option in UrQMD v4.0, which identifies reconstructable resonances based on the decay history of their daughter particles, i.e., whether they undergo further interactions after resonance decay. This procedure provides a proxy for the experimentally reconstructed signal based on the invariant-mass method. While the red points correspond to the total number of resonances that have decayed during the transport evolution.

A clear peak structure is observed for all three resonances, demonstrating that a substantial fraction of resonances can be reconstructed despite the presence of hadronic rescattering. In more central collisions, a stronger suppression of the reconstructed signal relative to the true distribution is observed, reflecting the increased probability of daughter-particle rescattering in a denser hadronic medium. In contrast, peripheral collisions exhibit a larger reconstructed fraction due to reduced hadronic interaction effects. The comparison between reconstructed and true distributions therefore provides a direct visualization of the medium modification of short-lived resonances in heavy-ion collisions.

\begin{figure}[htbp]
    \centering
    \includegraphics[width=0.85\columnwidth]{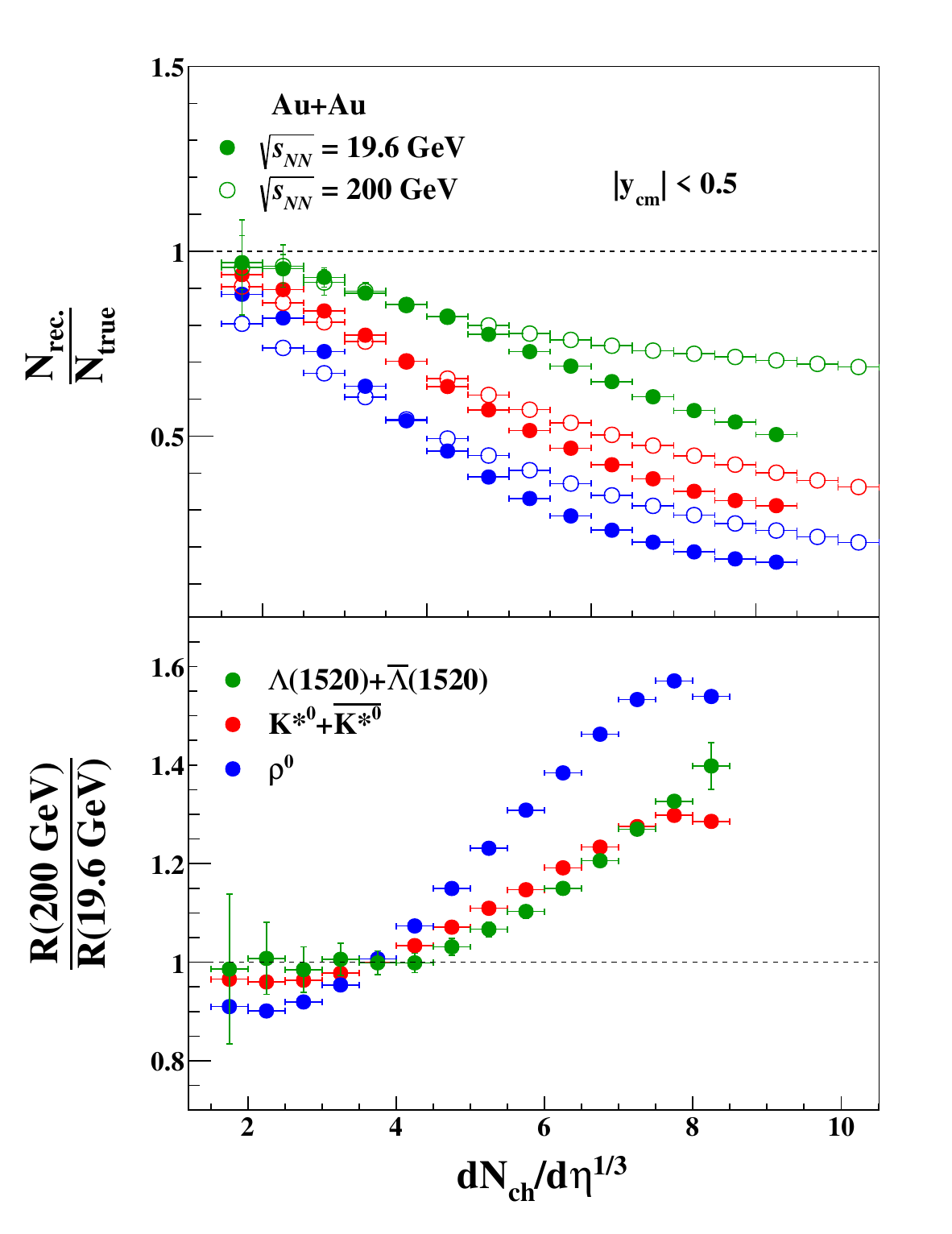}
    \caption{Reconstructable fraction $N_{\mathrm{rec}}/N_{\mathrm{true}}$ of short-lived resonances as a function of system size, characterized by $(dN_{ch}/d\eta)^{1/3}$, in Au+Au collisions at $\sqrt{s_{NN}} = 19.6 ~\mathrm{GeV}$ (left panel) and $200 ~\mathrm{GeV}$ (right panel) from UrQMD. 
    The upper panel shows the reconstructable fraction for for $K^{*0}+\overline{K}^{*0}$ (red), $\rho^0$ (blue), and $\Lambda(1520)+\overline{\Lambda}(1520)$ (green), with filled and open markers representing 19.6 and 200 GeV, respectively. The lower panel presents the ratio of the reconstructable fractions between the two collision energies, $R(200~\mathrm{GeV})/R(19.6~\mathrm{GeV})$. The dashed line at unity in the upper panel corresponds to the case where all decayed resonances are reconstructable, while that in the lower panel indicates no energy dependence.
    A clear hierarchy in reconstructability is observed, reflecting the different lifetimes and sensitivities of the resonances to hadronic interactions.}
    \label{fig:fig3_fraction_Nch13}
\end{figure}

Turn to the system-size dependence of the reconstructable fraction of short-lived resonances, as shown in Fig.~\ref{fig:fig3_fraction_Nch13} for Au+Au collisions at $\sqrt{s_{NN}} = 19.6$ and $200 ~\mathrm{GeV}$. The reconstructable fraction $N_{\mathrm{rec}}/N_{\mathrm{true}}$ is presented as a function of $(dN_{ch}/d\eta)^{1/3}$, which reflects the increasing system size and medium density.

A clear and universal suppression pattern is observed for all studied resonances, characterized by a monotonic decrease of the reconstructable fraction with increasing system size. This behavior reflects the increasing role of daughter-particle rescattering in a denser and longer-lived hadronic medium formed in more central collisions.

Among the three resonances, a clear hierarchy in the suppression of the reconstructable fraction is observed. The $\rho^0$ meson exhibits the strongest suppression, consistent with its short lifetime and high probability of decaying within the hadronic phase, where its decay daughters are susceptible to rescattering. The $K^{*0}$ shows an intermediate level of suppression, while the $\Lambda(1520)$, due to its relatively longer lifetime, exhibits the weakest suppression and correspondingly the largest reconstructable fraction across all centralities.

Comparing the two collision energies, similar qualitative trends are observed at both $\sqrt{s_{NN}} = 19.6$ and $200~\mathrm{GeV}$. 
The reconstructable fraction exhibit consistent dependences on the system size, with stronger suppression observed for shorter-lived resonances.
The energy dependence is relatively modest but non-negligible, as indicated by the ratio of reconstructable fractions between the two energies.
These differences may reflect variations in the space-time evolution of the hadronic phase.

These results provide a unified picture in which the reconstructability of short-lived resonances is primarily governed by their intrinsic lifetimes, while additional modifications arise from the space–time evolution of the hadronic phase.

\begin{figure}[htbp]
    \centering
    \includegraphics[width=1\columnwidth]{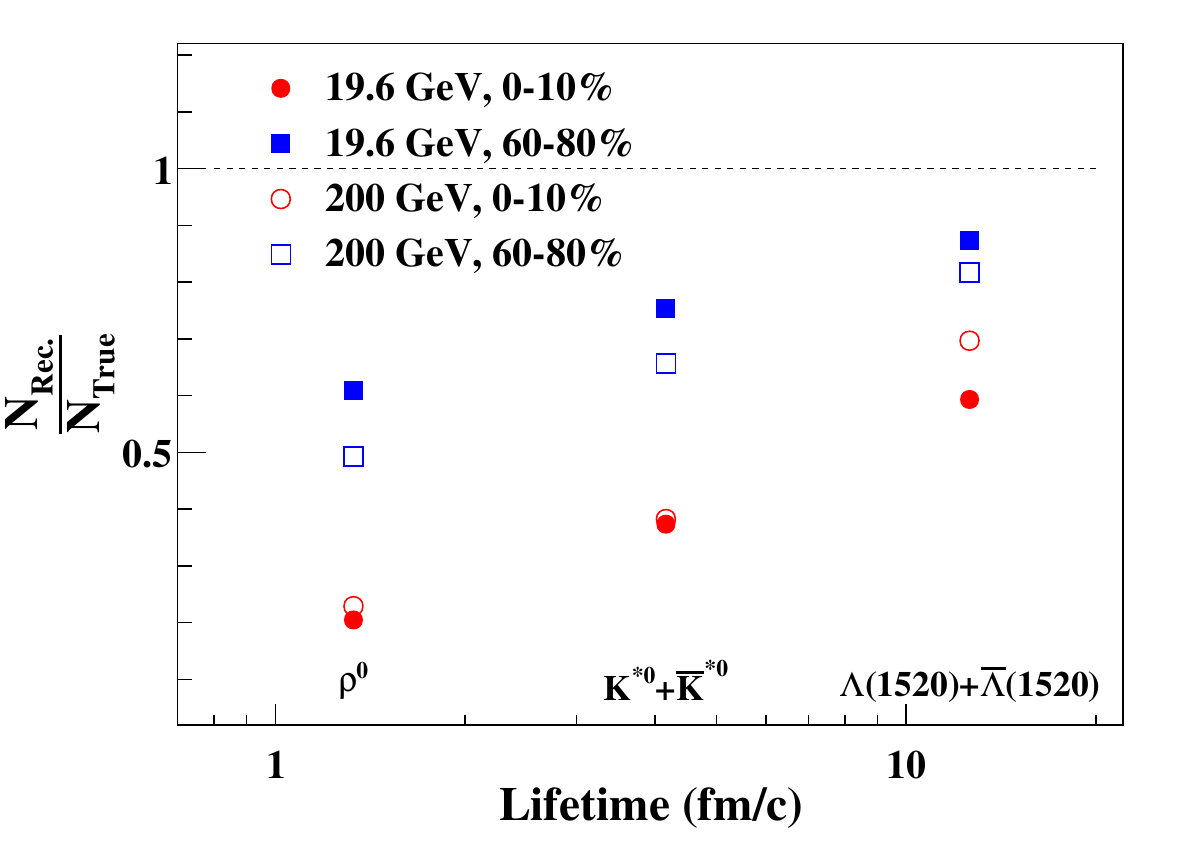}
    \caption{Reconstructable fraction $N_{\mathrm{rec}}/N_{\mathrm{true}}$ of short-lived resonances as a function of resonance lifetime in Au+Au collisions at $\sqrt{s_{NN}} = 19.6$ and $200 ~\mathrm{GeV}$ from UrQMD simulations. Results are shown for $\rho^0$, $K^{*0}+\overline{K}^{*0}$, and $\Lambda(1520)+\overline{\Lambda}(1520)$, corresponding to different lifetimes on the x-axis. Filled symbols represent 19.6 GeV results, while open symbols correspond to 200 GeV. Red and blue markers denote $0–10\%$ and $60–80\%$ centrality classes, respectively. The dashed line at unity corresponds to the case where all decayed resonances are reconstructable. A clear lifetime ordering of reconstructability is observed, demonstrating stronger suppression for shorter-lived resonances and enhanced medium effects in more central collisions.}
    \label{fig:fig4_fraction_lifetime}
\end{figure}

Then we present a comprehensive summary of the resonance reconstructability as a function of resonance lifetime in Au+Au collisions at $\sqrt{s_{NN}} = 19.6$ and $200 ~\mathrm{GeV}$, as shown in Fig.~\ref{fig:fig4_fraction_lifetime}. The results are shown for $\rho^0$, $K^{*0}+\overline{K}^{*0}$, and $\Lambda(1520)+\overline{\Lambda}(1520)$, covering a wide range of lifetimes and allowing for a direct study of lifetime-dependent medium effects.

This ordering is consistently observed as a function of resonance lifetime, indicating that the suppression pattern is primarily governed by the intrinsic lifetimes of the resonances. Shorter-lived states are more sensitive to hadronic rescattering effects, leading to a reduced reconstructable fraction in more central collisions and at higher system sizes.

In addition to the lifetime dependence, a clear centrality dependence is observed, where more central collisions show stronger suppression compared to peripheral collisions. This behavior is consistent with the picture illustrated in Fig.~\ref{fig:fig2_invmass}, where hadronic rescattering leads to a reduction of the reconstructed resonance signal. In more central collisions, the increased hadronic medium density and longer interaction time further enhance the probability of daughter rescattering, resulting in a stronger suppression.

Comparing the two collision energies, similar qualitative behavior is observed at both $\sqrt{s_{NN}} = 19.6$ and $200 ~\mathrm{GeV}$, while quantitative differences suggest variations in the effective hadronic phase properties, such as lifetime, density, and baryon composition.

Overall, the results demonstrate a unified picture in which the reconstructability of short-lived resonances is primarily governed by their intrinsic lifetimes, while additional modifications arise from the system size and collision energy dependence of the hadronic phase.

\subsection{Directed flow}
\begin{figure}[htbp]
    \centering
    \includegraphics[width=1\columnwidth]{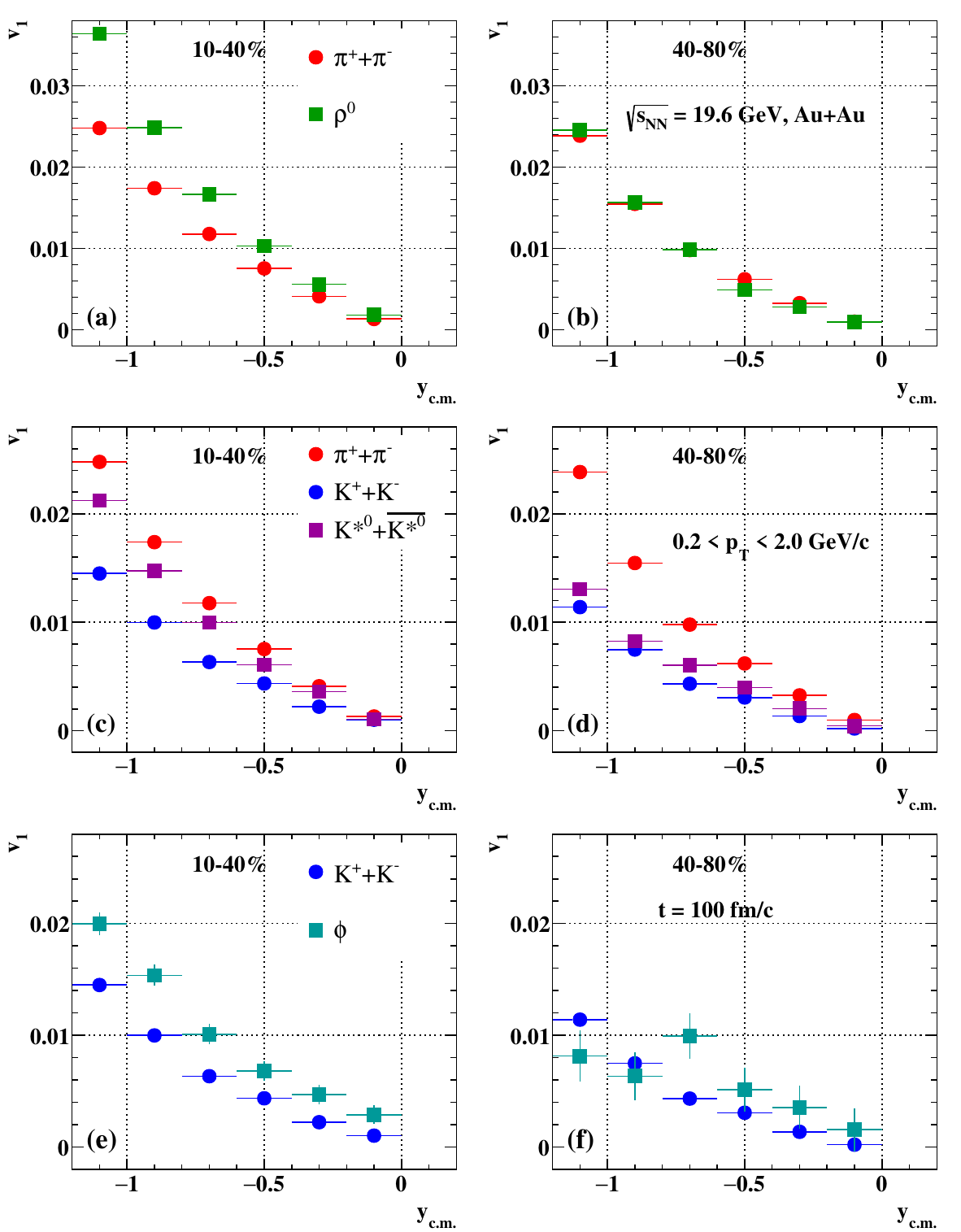}
    \caption{Directed flow $v_1$ as a function of center-of-mass rapidity $y_{c.m.}$ for different particle species in Au+Au collisions at $\sqrt{s_{NN}} = 19.6~\mathrm{GeV}$. Results are shown for $\pi^+ + \pi^-$ (red circles), $K^+ + K^-$ (blue circles), $\rho^0$ (green squares), $K^{*0} + \bar{K}^{*0}$ (purple squares), and $\phi$ (cyan squares). The left (right) panels correspond to $10–40\%$ ($40–80\%$) centrality classes. The upper, middle, and lower panels show comparisons between $\pi^+ + \pi^-$ and $\rho^0$, $\pi/K$ and $K^{*0}$, and $K$ and $\phi$, respectively.}
    \label{fig:fig5_v1_y_19p6GeV}
\end{figure}

After discussing the reconstructability of short-lived resonances, we investigate the directed flow of resonances during the hadronic evolution.
Fig.~\ref{fig:fig5_v1_y_19p6GeV} shows the rapidity dependence of $v_1$ for $\pi^+ + \pi^-$, $K^+ + K^-$, $\rho^0$, $K^{*0}+\overline{K}^{*0}$ and $\phi$ in Au+Au collisions at $\sqrt{s_{NN}} = 19.6~\mathrm{GeV}$, for two representative centrality classes, $10-40\%$ and $40-80\%$.

In mid-central collisions ($10–40\%$), short-lived resonances exhibit a larger directed flow magnitude than stable mesons, with the $\rho^0$ showing the largest $v_1$ slope. 
The difference between pion and kaon $v_1$ reflects their different responses to the initial dynamics and interactions with the baryon-rich hadronic medium during the system evolution. 
The observed differences between resonances and stable mesons suggest that their final-state $v_1$ values probe different stages of the system evolution.
At late evolution times, the reconstructed samples of short-lived resonances such as $\rho^0$ and $K^{*0}$ are dominated by resonances regenerated during the hadronic phase, and therefore mainly reflect the collective motion of the hadronic system at the time of their last formation.
In contrast, stable mesons retain contributions from different stages of the evolution through their initial production and subsequent interactions with the hadronic medium.

However, the observed resonance $v_1$ modification does not follow a simple lifetime ordering. In particular, despite the significantly different lifetimes of the $K^{*0}$ ($\tau\sim4 fm/c$) and the $\phi$ ($\tau\sim46 fm/c$), they exhibit comparable $v_1$ values at backward rapidity, while the $\phi$ shows a larger $v_1$ magnitude than both kaons and $K^{*0}$ at mid-rapidity. These observations indicate that resonance lifetime alone is insufficient to determine the final directed flow.

In contrast, in peripheral collisions ($40-80\%$), the shorter duration and lower density of the hadronic phase reduce both hadronic rescattering and resonance regeneration. Consequently, the late-stage hadronic evolution has a much weaker impact on the final-stage $v_1$.
Stable mesons retain directed flow values closer to those established at earlier stages, while reconstructed short-lived resonances receive a smaller contribution from late-stage regeneration and preserve a larger fraction of the information from their earlier production. As a result, the differences in $v_1$ between short-lived resonances and stable hadrons become much smaller.

Specifically, the $\rho^0$  $v_1$ is very close to that of pions, while the $K^{*0}$ $v_1$ is closer to kaons than to pions at both mid-rapidity and backward rapidity. In contrast, the difference between the $\phi$ $v_1$ and that of kaons at mid-rapidity remains larger than the corresponding difference observed for $K^{*0}$. This behavior cannot be explained by the resonance lifetime alone, suggesting that additional mechanisms may contribute to the final directed flow. A more detailed understanding of the underlying dynamics requires further investigation.

\begin{figure}[htbp]
    \centering
    \includegraphics[width=1\columnwidth]{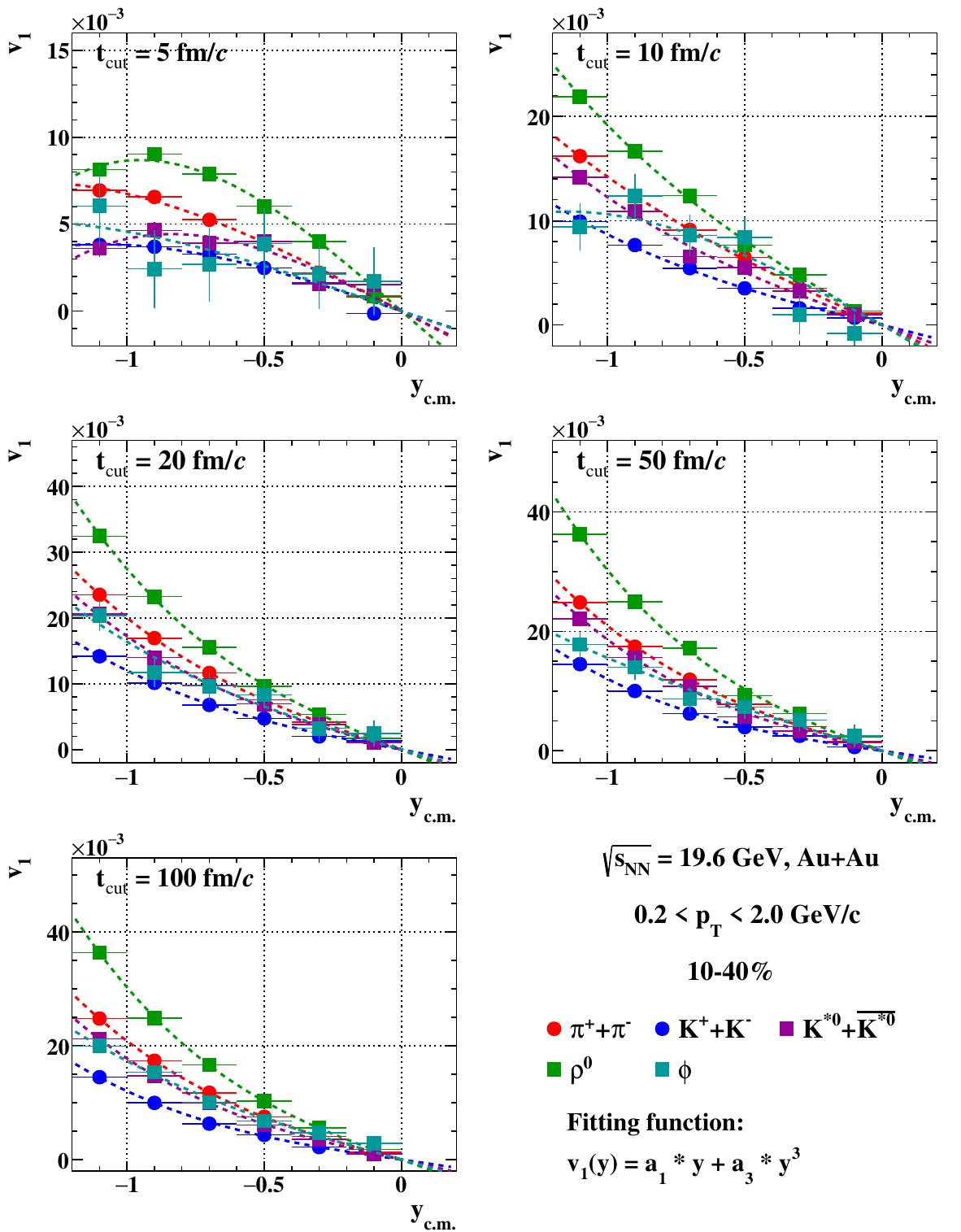}
    \caption{Time evolution of the rapidity dependence of directed flow $v_1$ for $\pi^+ + \pi^-$, $K^+ + K^-$, $\rho^0$, $K^{*0} + \bar{K}^{*0}$, and $\phi$ in Au+Au collisions at $\sqrt{s_{NN}} = 19.6~\mathrm{GeV}$ for $10–40\%$ centrality. The results are shown at different evolution cutoff times $t_{\mathrm{cut}}$ = 5, 10, 20, 50, and 100$~\mathrm{fm}/c$. The directed flow if fitted with the function $v_1(y)=a_1*y+a_3*y^3$.}
    \label{fig:fig6_v1_y_time_19p6GeV}
\end{figure}

\begin{figure}[htbp]
    \centering
    \includegraphics[width=0.95\columnwidth]{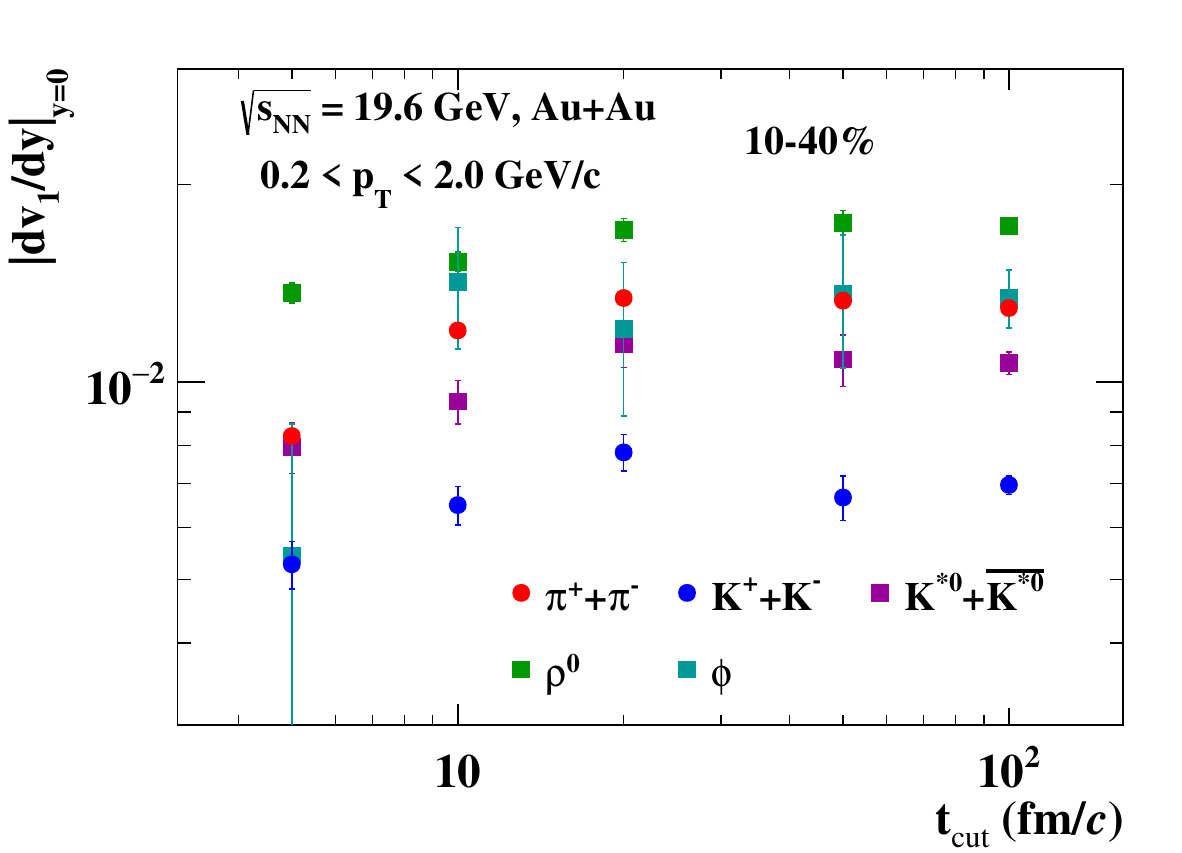}
    \caption{Time evolution of the absolute value of the midrapidity slope $|dv_1/dy|_{y=0}$ for $\pi^+ + \pi^-$, $K^+ + K^-$, $\rho^0$, $K^{*0} + \bar{K}^{*0}$, and $\phi$ in Au+Au collisions at $\sqrt{s_{NN}} = 19.6~\mathrm{GeV}$ for $10–40\%$ centrality. The results are shown at different evolution cutoff times $t_{\mathrm{cut}}$ = 5, 10, 20, 50, and 100$~\mathrm{fm}/c$.}
    \label{fig:fig7_v1_slope_time_19p6GeV}
\end{figure}

To further understand the dynamical origin of the observed species dependence, we study the time evolution of the directed flow for different particle species in $10–40\%$ Au+Au collisions at $\sqrt{s_{NN}} = 19.6~\mathrm{GeV}$. Fig.~\ref{fig:fig6_v1_y_time_19p6GeV} shows the rapidity dependence of $v_1$ at different evolution cutoff times $t_{\mathrm{cut}}$ = 5, 10, 20, 50, and 100~$\mathrm{fm}/c$. The rapidity dependence is described using the function $v_1(y)=a_1y+a_3y^3$, from which the midrapidity slope is extracted and discussed in Fig.~\ref{fig:fig7_v1_slope_time_19p6GeV}.

These snapshots provide a direct view of how the late-stage hadronic evolution modifies the directed flow of different particle species. Although all particles exhibit a negative $v_1$ slope, the evolution of its magnitude differs substantially among different particle species.

At t = 5~$\mathrm{fm}/c$, the $\rho^0$ already exhibits a noticeable deviation from the pion $v_1$, whereas the $K^{*0}$ and $\phi$ remain much closer to the kaon $v_1$. 
The early deviation of the $\rho^0$ $v_1$ is consistent with its short lifetime, making it more sensitive to the hadronic evolution at early times than $K^{*0}$ and $\phi$.
As the system evolves from $t_{\mathrm{cut}}$ = 20 to 100~$\mathrm{fm}/c$, the difference between resonances and their corresponding stable mesons are largely established and remain nearly unchanged throughout the subsequent evolution.

The evolution of the extracted $|dv_1/dy|_{y=0}$ shown in Fig.~\ref{fig:fig7_v1_slope_time_19p6GeV} further quantifies these observations. The absolute values of the $v_1$ slopes increase rapidly from $t_{\mathrm{cut}}$ = 5 to 20~$\mathrm{fm}/c$, indicating that the major development of directed flow occurs within $0 \sim 20~\mathrm{fm}/c$ of the hadronic evolution.
Beyond $t_{\mathrm{cut}} \approx 20~\mathrm{fm}/c$, the slopes become nearly constant, indicating that the hadronic modification of the directed flow is largely established by this time. The relative ordering among different particles species also remains nearly unchanged during the subsequent evolution.
Interestingly, the $\phi$ develops a $v_1$ slope comparable to that of pions at late evolution times despite its much longer lifetime. The origin of this behavior is not yet fully understood and deserves further investigation.

\section{Summary}

We have performed a systematic study of both the reconstructability and the directed flow of short-lived hadronic resonances in Au+Au collisions at $\sqrt{s_{NN}}=19.6$ and $200~\mathrm{GeV}$ using the UrQMD transport model. The reconstructability was investigated through comparisons between truth-level and reconstructable resonance yields, while the directed flow was studied to explore the influence of hadronic evolution on the final-state $v_1$.

Three representative resonances, $\rho^0$, $K^{*0}$, and $\Lambda(1520)$, were investigated to study reconstructability over a broad range of lifetimes and decay channels. Their reconstructable fractions were analyzed as functions of collision centrality, system size, and resonance lifetime.

The results show a clear hierarchy in reconstructability, where shorter-lived resonances exhibit stronger suppression due to increased daughter rescattering in the hadronic medium. This behavior is consistently observed at both collision energies and across all centrality classes. A pronounced centrality dependence is also found, with stronger suppression in central collisions, reflecting the larger size and longer lifetime of the hadronic medium. Furthermore, a non-trivial collision-energy dependence is observed between $\sqrt{s_{NN}}=19.6$ and $200~\mathrm{GeV}$, indicating that the properties of the hadronic medium also affect resonance reconstructability.

The directed-flow analysis reveals a clear differences between resonances and their corresponding stable hadrons in mid-central collisions, while these differences become significantly weaker in peripheral collisions. The time evolution of the $v_1$ further demonstrates that different resonances species exhibit distinct evolution patterns during the hadronic stage, indicating that hadronic evolution plays an important role in shaping the final-state $v_1$.

Overall, the present study provides a systematic understanding of how the hadronic phase influences both resonance reconstructability and directed flow in relativistic heavy-ion collisions.
These findings provide useful guidance for the interpretation of experimental measurements of short-lived resonances, such as $\rho^0$, $K^{*0}$, and $\phi$, and help to disentangle the effects of hadronic evolution from the observed resonance yields and collective flow.

\FloatBarrier
\begin{acknowledgments}
This work is supported in part by the National Natural Science Foundation of China under Grant No. 12375134, the National Key Research and Development Program of China (Grant No. 2024YFE0110103 and 2024YFA1611003), and the Fundamental Research Funds for the Central Universities (Grant No. CCNU25JCPT017).
\end{acknowledgments}


\bibliography{main}

\appendix

\end{document}